\documentclass[aps,prx,twocolumn,superscriptaddress]{revtex4-2}

\usepackage[T1]{fontenc}
\usepackage{graphicx}
\usepackage{dcolumn}
\usepackage{bm}
\usepackage{amsmath}
\usepackage{braket}
\usepackage[english]{babel}
\usepackage[utf8]{inputenc}
\usepackage[dvipsnames]{xcolor}
\usepackage{soul}
\usepackage{units}
\colorlet{RED}{red}

\usepackage{hyperref}
\allowdisplaybreaks

\begin{document}

\title{Noise-protected two-qubit gate using anisotropic exchange interaction}
\author{Zizheng Wu}
\affiliation{QuTech and Kavli Institute of Nanoscience, Delft University of Technology, Delft, The Netherlands}
\author{Maximilian Rimbach-Russ}
\email{ m.f.russ@tudelft.nl}
\affiliation{QuTech and Kavli Institute of Nanoscience, Delft University of Technology, Delft, The Netherlands}

\begin{abstract}

Hole spin qubits hosted in Germanium quantum dots are promising candidates for scalable quantum computing. The strong spin-orbit interaction can enable fast and all-electrical quantum control. Furthermore, the platform can implement universal quantum control using only baseband signals, which may mitigate the impact of crosstalk and microwave-induced heating. At the same time, spin-orbit interaction gives rise to an anisotropic exchange interaction, whose potential for implementing two-qubit gates has remained largely unexplored. However, the current performance of operating a hole-based quantum computer is mostly limited by dephasing due to low-frequency charge noise. In this work, we propose a novel two-qubit gate protocol for Germanium hole spin qubits operated in the gapless regime. This gate protocol exploits the anisotropic exchange interaction between neighboring spins and utilizes a composite pulse scheme implemented solely through electrical baseband signals. Using this approach, we predict high-fidelity two-qubit controlled-Z operations that can suppress exchange-energy fluctuations, offering a pathway toward fault-tolerant semiconductor quantum processors.
\end{abstract}

\date{\today}

\maketitle

\section{Introduction}
\label{Section: Introduction}
Semiconductor quantum dot spin qubits have emerged as a highly promising platform for quantum computing owing to their excellent compatibility with well-established semiconductor fabrication technologies~\cite{burkardSemiconductorSpinQubits2023,vandersypenInterfacingSpinQubits2017,neyensProbingSingleElectrons2024,steinacker300MmFoundry2024,huckemannIndustriallyFabricatedSingleelectron2024}. In particular, Germanium (Ge) hole spin qubits are highly appealing, as their strong spin-orbit interaction (SOI) enables ultrafast all-electric qubit manipulation~\cite{fangRecentAdvancesHolespin2023,froningUltrafastHoleSpin2021,wangUltrafastCoherentControl2022,johnBichromaticRabiControl2024,wangOperatingSemiconductorQuantum2024a,burkardSemiconductorSpinQubits2023,stanoReviewPerformanceMetrics2022,vandersypenInterfacingSpinQubits2017,philipsUniversalControlSixqubit2022,xueQuantumLogicSpin2022,millsTwoqubitSiliconQuantum2022,zwerver2023shuttling,de2025high,takedaQuantumErrorCorrection2022,noiri2022shuttling,steinacker2025industry,gonzalez-zalbaScalingSiliconbasedQuantum2021,petitUniversalQuantumLogic2020,siegel2025snakes,hetenyiTailoringQuantumError2024,svastits2025readout,liFlexible300Mm2020,kloeffelProspectsSpinBasedQuantum2013,maurandCMOSSiliconSpin2016,hendrickxSweetspotOperationGermanium2024,tidjani2025three}. So far, high-fidelity quantum gates have already been demonstrated experimentally, with single-qubit gates fidelity exceeding $99.9\%$~\cite{lawrieSimultaneousSinglequbitDriving2023,zhou2025high} and two-qubit gates reaching $99\%$~\cite{wangOperatingSemiconductorQuantum2024a}. Alongside these advances, baseband control techniques, such as spin hopping, have also achieved significant progress in enabling high-fidelity qubit manipulation in 2D Ge quantum-dot arrays~\cite{vanriggelen-doelmanCoherentSpinQubit2024,wangOperatingSemiconductorQuantum2024a} and suppressing crosstalk and heating effects induced by microwave signals~\cite{undsethNonlinearResponseCrosstalk2023, undsethHotterEasierUnexpected2023}.

An important feature of SOI is that it can lead to an anisotropic exchange interaction~\cite{hetenyiExchangeInteractionHolespin2020, saez-mollejoMicrowaveDrivenSinglettriplet2024, geyerAnisotropicExchangeInteraction2024}. Since the development of electron spin qubits preceded that of hole spin qubits, at present, most efforts in modeling, control, and optimization of exchange interactions are mainly concentrated on the isotropic coupling between electron spins~\cite{russHighfidelityQuantumGates2018,rimbach-russSimpleFrameworkSystematic2023}. Strategies that fully exploit the anisotropic exchange interaction between hole spin qubits for realizing high-fidelity two-qubit gates are still in their early stages of development. Nevertheless, it has been predicted that the SOI-induced anisotropy can be harnessed to implement controlled-NOT (CNOT) gates~\cite{geyerAnisotropicExchangeInteraction2024}, and recent experimental efforts have begun exploring its use for realizing iSWAP gates~\cite{tsoukalas2025resonant}. Analogous to the isotropic exchange interaction between electron spin qubits, anisotropic exchange is also susceptible to omnipresent charge noise in semiconductors~\cite{dialChargeNoiseSpectroscopy2013,martinsNoiseSuppressionUsing2016,burkardSemiconductorSpinQubits2023,paladinoNoiseImplicationsSolidstate2014,paquelet2023reducing,lodariLowPercolationDensity2021,wangOperatingSemiconductorQuantum2024a,stehouwer2025exploiting,hendrickxSweetspotOperationGermanium2024,piotSingleHoleSpin2022}. To overcome this limitation, composite-pulse schemes~\cite{vandersypenNMRTechniquesQuantum2005,gungorduPulseSequenceDesigned2018} have been proposed as an effective means of suppressing low-frequncy charge noise.

In this work, we propose an innovative approach that harnesses the anisotropic exchange interaction between Ge hole spin qubits to implement a two-qubit controlled-Z (CZ) gate. This CZ gate is realized using a specific composite-pulse scheme known as Short Composite Rotation For Undoing Length Over- and Undershoot (SCROFULOUS)~\cite{cumminsTacklingSystematicErrors2003,ichikawa2013minimal}. Compared with conventional adiabatic CZ gates~\cite{gungorduPulseSequenceDesigned2018}, our proposed gate protocol enables faster operation times and is realized entirely through electrical baseband signals without the need of microwaves. Moreover, this composite-pulse gate effectively suppresses the detrimental impact of exchange-energy fluctuations on gate performance, thereby enhancing overall fidelity and noise resilience. 

The structure of this paper is organized as follows. Section~\ref{Section: Qubit model and Hamiltonian} and~\ref{Section: Composite poulse scheme} lay the theoretical foundation of our work. In Sec.~\ref{Section: Qubit model and Hamiltonian}, we introduce the anisotropy of planar Ge holes' g-tensor and present the Hamiltonian model describing two Ge hole spin qubits hosted in a double quantum dot (DQD) system. Within this framework, we explain the implementation of a ZZ phase rotation and introduce the noise model. In Sec.~\ref{Section: Composite poulse scheme}, we briefly introduce the composite pulse scheme. Subsequently, in Sec.~\ref{Section: Noise-protected CZ gate}, we elaborate on the implementation of a two-qubit CZ gate that is protected against quasi-static exchange-interaction fluctuations, analyze relevant experimental error sources, characterize the gate's robustness against spectral noise and present the main simulation results. In Sec.~\ref{Section: Conclusion and Outlook} we conclude the paper and discuss future directions.

\begin{figure*}[t]
    \centering
    \includegraphics[width=0.9\textwidth]{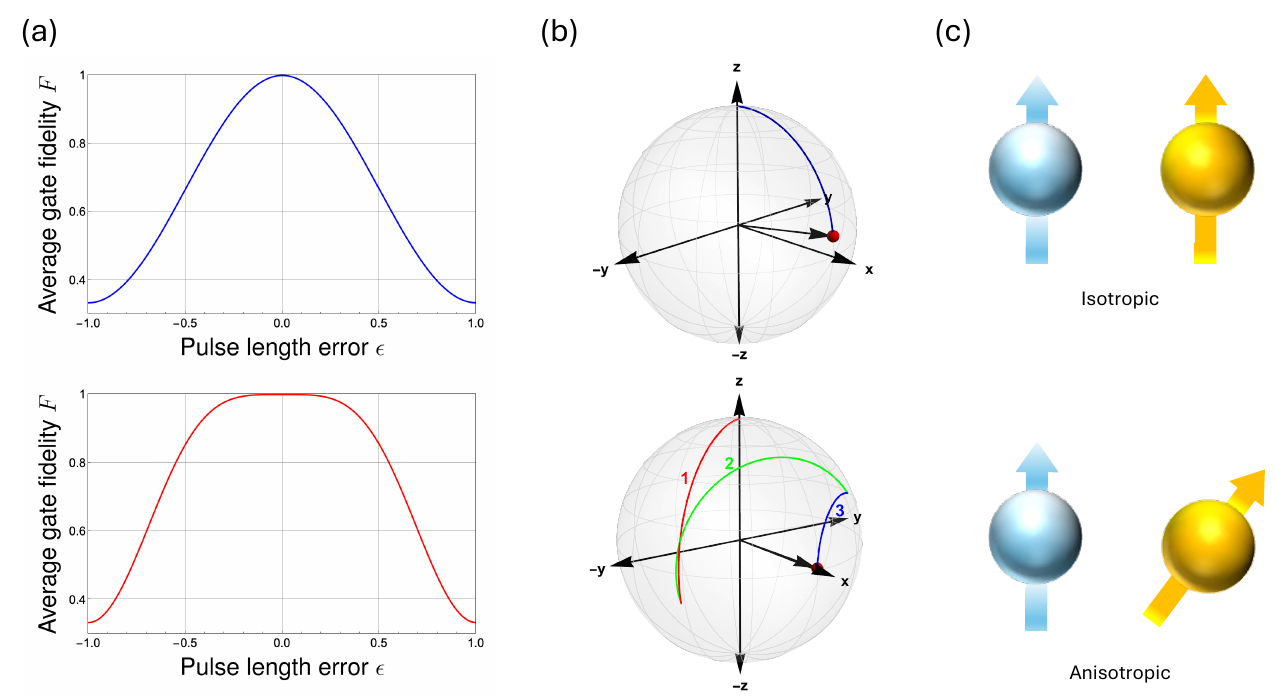}
    \caption{(a) Simulated average gate fidelities (Eq.~\ref{Average gate fidelity}) of a single-pulse X gate (top) and a SCROFULOUS composite-pulse X gate (bottom) in the presence of pulse length errors. (b) Bloch-sphere trajectories for a single-pulse (top) and a SCROFULOUS composite-pulse (bottom) $R_y(\pi/2)$ gate. Each individual rotation is assumed to experience a $10\%$ under-rotation error. Starting from the $|0\rangle$ state (pointing along $+z$), the final state after the single-pulse gate deviates significantly from the target position along the $+x$ direction, whereas the SCROFULOUS gate brings the state much closer to the target. (c) Illustration of isotropic and anisotropic exchange interactions between spins in adjacent quantum dots. The isotropic exchange is described by $H = J_0\,\vec S_1 \cdot \vec S_2$, whereas the anisotropic exchange takes the form $H = J_0\,\vec S_1 \cdot (R\,\vec S_2)$, with $R$ a rotation matrix.}
    \label{fig:figure_1}
\end{figure*}

\section{Qubit Model and Hamiltonian}
\label{Section: Qubit model and Hamiltonian}
\subsection{The anisotropy of g-tensor}
\label{Section: The anisotropy of g-tensor}
A single hole state inside a gate defined quantum dot can be well-described by the Luttinger-Kohn Hamiltonian supplemented by the Bir-Pikus Hamiltonian which accounts for the strain effects~\cite{roland2003spin,willatzen2009kp}
\begin{align}
    \mathcal{H}=\mathcal{H}_{\text{LK}}(\vec{{k}})+\mathcal{H}_{\text{BP}}.
    \label{ham:HLKBP}
\end{align}
The effective low-energy dynamics can be obtained by projecting onto the ground-state spin-$3/2$ subspace. In a planar Ge heterostructure, confinement and/or strain further induce a large heavy-hole/light-hole splitting $\Delta_{\mathrm{HL}}$, which separates the $J_z=\pm 3/2$ states from the $J_z=\pm 1/2$ states and results in a well-isolated Kramers pair that defines the qubit. The interaction between the qubit and the magnetic field can be written as
$H_q = \frac{1}{2} \mu_B\vec{B}\cdot \mathcal{G}\vec{\sigma}$~\footnote{Another commonly used form is $H_q=\frac{1}{2}\mu_B\vec{\sigma}\cdot\mathcal{G}'\vec{B}$, 
where the two representations are related by $\mathcal{G}'=\mathcal{G}^{\mathsf{T}}$.},
where $\mu_B$ is Bohr's magneton, $\vec{B}$ is the magnetic field vector, and $\mathcal{G}$ is the effective g-tensor which arises from the linear and cubic Zeeman terms, as well as the generalized momentum $\vec{k}\rightarrow\vec{k}+eA$. Throughout this work, we adopt the unit where $\hbar = 1$. The in-plane $\mathcal{G}$ in gate-defined quantum dots hosted in planar Ge heterostructure is highly tunable through electric fields, as reflected in the expressions of the g-tensor elements given below. Assuming directionally separable wavefunctions, such that $\langle p_ip_j \rangle=0$ for $i\neq j$, the g-tensor components can be described by the following approximate expressions~\cite{rimbach2025gapless,abadillo-urielHoleSpinDrivingStrainInduced2023,venitucciSimpleModelElectrical2019,michalLongitudinalTransverseElectric2021,rodriguez2025sweet,valvo2025electrically}
\begin{subequations}\label{eq:g_tensor}
\begin{align}
    g_{xx} &= 3 q 
    - \frac{6 \tilde{\kappa} b_v \bigl( \langle \varepsilon_{xx} \rangle 
    - \langle \varepsilon_{yy} \rangle \bigr)}{\Delta_{\mathrm{HL}}}
    - \frac{6 \bigl( \lambda \langle p_x^2 \rangle 
    - \lambda' \langle p_y^2 \rangle \bigr)}
    {m_0 \Delta_{\mathrm{HL}}}, \label{eq:g_xx} \\
    g_{yy} &= -3 q 
    - \frac{6 \tilde{\kappa} b_v \bigl( \langle \varepsilon_{xx} \rangle 
    - \langle \varepsilon_{yy} \rangle \bigr)}{\Delta_{\mathrm{HL}}}
    + \frac{6 \bigl( \lambda \langle p_y^2 \rangle 
    - \lambda' \langle p_x^2 \rangle \bigr)}
    {m_0 \Delta_{\mathrm{HL}}}, \label{eq:g_yy} \\
    g_{zz} &= 6\kappa + \frac{27}{2} q - 2\gamma_h, 
    \label{eq:g_zz} \\
    g_{xy} &= - g_{yx}
    = - \frac{4 \sqrt{3} \kappa d_v \varepsilon_{xy}}
    {\Delta_{\mathrm{HL}}}, \label{eq:g_xy} \\
    g_{xz} &= - g_{zx}
    = - \frac{4 \sqrt{3} \kappa d_v \varepsilon_{xz}}
    {\Delta_{\mathrm{HL}}}, \label{eq:g_xz} \\
    g_{yz} &= - g_{zy}
    = - \frac{4 \sqrt{3} \kappa d_v \varepsilon_{yz}}
    {\Delta_{\mathrm{HL}}}. \label{eq:g_yz}
\end{align}
\end{subequations}
Here, $\lambda$, $\lambda'$ and $\tilde{\kappa}$ are material-dependent scaling parameters given by $\lambda = 2 \eta_h \gamma_3^2 - \tilde{\kappa} \gamma_2$, $\tilde{\lambda} = 2  \eta_h (\gamma_2 \gamma_3 + \gamma_3^2) - \tilde{\kappa} \gamma_3$,
and $\tilde{\kappa} = \kappa - 2 \tilde{\eta}_h \gamma_3$. The parameters $\gamma_2$, $\gamma_3$, $\kappa$, and $q$ are the Luttinger parameters, while $\eta_h$ and $\tilde{\eta}_h$ arise from interband coupling. In this work, we adopt the values $\gamma_2=4.24$, $\gamma_3=5.69$, $\kappa=3.41$, $q=0.06$ and $\gamma_h=2.62$~\cite{abadillo-urielHoleSpinDrivingStrainInduced2023}. The free electron mass is given by $m_0$ and $d_v$ represents the shear deformation potential of the valence band. The choice of $\eta_h$ and $\tilde{\eta}_h$ will be discussed in Appendix~\ref{Appendix C}. By further assuming negligible shear strain $\varepsilon_{xy}=\varepsilon_{yz}=\varepsilon_{xz}=0$, the effective g-tensor becomes diagonal. From the expression, we can see that $\langle \varepsilon_{xx} \rangle - \langle \varepsilon_{yy} \rangle$ only lead to a shift in the value of $g_{xx}$ and $g_{yy}$. We therefore neglect their contribution, as their effect can be compensated by adjusting the size of the quantum dot~\cite{rimbach2025gapless,valvo2025electrically}. Finally, we emphasize that $\langle p_i^2\rangle, i\in \{x,y\}$ can be tuned electrically. 

\subsection{DQD system}
To describe the two-qubit dynamics inside a  DQD system filled with two holes, we use a Fermi-Hubbard model with a single accessible orbital per site. Mapping it onto a DQD system gives rise to
\begin{align}
    \mathcal{H}_{\text{DQD}}
    &=
    \sum_{ij\in\{1,2\}}\sum_{ss'\in\{\uparrow,\downarrow\}} 
    a^{\dagger}_{i s} \bar{\mathcal{H}} a_{j s'} + U\sum_{i\in\{1,2\}} n_{i\uparrow} n_{i\downarrow},
    \label{eq:DQD Hamiltonian}
\end{align}
where $a^{\dagger}_{i s}$ ($a_{i s}$) creates (annihilates) a hole on site $i$ with spin $s$ and satisfies the fermionic anticommutation relations. The operator $n_{i s} = a^{\dagger}_{i s} a_{i s}$ denotes the particle number with spin $s$ on dot $i$, $U$ represents the effective charging energy, and $\bar{\mathcal{H}}$ is the single-particle Hamiltonian~\cite{geyerAnisotropicExchangeInteraction2024}
\begin{align}
    \bar{\mathcal{H}}=
    &\frac{1}{2}\mu_B\mathbf{B}\cdot
    \left( \frac{1 + \tau_z}{2}\,\hat{g}_1\boldsymbol{\sigma} + \frac{1 - \tau_z}{2}\,\hat{g}_2\boldsymbol{\sigma} \right) + \frac{\epsilon_{\text{eff}}}{2}\tau_z \notag \\ 
    & + t\cos(\phi_{\text{so}})\tau_x + t\sin(\phi_{\text{so}})\tau_y\,\mathbf{n}_{\text{so}}\cdot\boldsymbol{\sigma}.
    \label{eq:single-particle Hamiltonian}
\end{align}
Here $\tau_{i},i\in{\{x,y,z\}}$ denote the Pauli operators acting on the orbital degrees of freedom, $\hat{g}_1$ and $\hat{g}_2$ are the corresponding g-tensors, and $(1\pm\tau_z)/2$ project on the orbital states $|1\rangle$ and $|2\rangle$ respectively. We denote by $\boldsymbol{\sigma}$ the Pauli vector, whose components are Pauli operators acting on the spin-$|\!\uparrow\rangle, |\!\downarrow\rangle$ space. The detailed expressions for $\tau$ and $\boldsymbol{\sigma}$ are given in Appendix~\ref{Appendix A}. The detuning between the two quantum dots is given by $\frac{\epsilon_{\text{eff}}}{2}\tau_z$. The last two terms  $t\cos(\phi_{\text{so}})\tau_x,t\sin(\phi_{\text{so}})\tau_y\,\mathbf{n}_{\text{so}}\cdot\boldsymbol{\sigma}$ describe the spin-conserving tunneling and spin-flip tunneling between the two quantum dots. Lastly, $t$ is the tunneling constant, $\phi_{\text{so}}$ is the rotation angle of the spin quantization axis during tunneling caused by the SOI, and $\mathbf{n}_{\text{so}}$ is the spin-orbit axis direction. 

We first simplify the Hamiltonian $\mathcal{H}_{\text{DQD}}$ by performing a unitary basis transformation defined by $U_{\text{so}}=\exp\left(-i\phi_{\text{so}}\tau_z\mathbf{n}_{\text{so}}\cdot\boldsymbol{\sigma}/2\right)$. This transformation rotates the spin-quantization axes in opposite directions at the two sites. Afterwards, we restrict ourselves to the $(1,1)$ charge configuration. Here $(n,m)$ denotes the charge configuration of the DQD system, with $n$ ($m$) holes in the left (right) quantum dot. By introducing effective spin operators, the resulting Hamiltonian can be written as a Heisenberg-like spin model, from which new frames can be derived via basis transformations. The detailed derivations are presented in Appendix~\ref{Appendix A}. In the following, we introduce two important frames used in our analysis.

\textit{Lab frame}—In the lab frame, $\mathcal{H}^{\text{lab}}$ reads
\begin{align}
    \mathcal{H}^{\text{lab}} = \frac{1}{2}\mu_B\mathbf{B}\cdot\hat{g}_1\boldsymbol{\sigma}_1
    + \frac{1}{2}\mu_B\mathbf{B}\cdot\hat{g}_2\boldsymbol{\sigma}_2 + \frac{1}{4}\boldsymbol{\sigma}_1\cdot \mathcal{J}\boldsymbol{\sigma}_2,
    \label{eq: lab-frame Hamiltonian}
\end{align}
where in this basis $\mathcal{J}=J_0\hat{R}_{\mathbf{n}_{\text{so}}}(-2\phi_{\text{so}})$ becomes a $3\times3$ real-valued tensor. Here $\hat{R}_{\mathbf{n}_{\text{so}}}(-2\phi_{\text{so}})$ is a rotation matrix around the spin-orbit vector $\mathbf{n}_{\text{so}}$ with spin-orbit angle $\phi_{\text{so}}$. This clearly shows that due to the SOI, the resulting exchange interaction is no longer isotropic and scalar but instead corresponds to a spin rotation (see Fig.~\ref{fig:figure_1}(c)). In this frame, $\hat{g}_1$, $\hat{g}_2$, and $\mathcal{J}$ can be directly expressed in terms of experimentally measured values, allowing the voltage dependence of both the $g$-tensor and the exchange coupling $\mathcal{J}$-tensor to be incorporated straightforwardly. Therefore, the lab frame is particularly suitable for describing the system Hamiltonian under electrical control.

\textit{Qubit frame}—Once the lab frame is obtained, we can further simplify the Zeeman term by applying an additional rotation matrix, thereby transforming the Hamiltonian into a new basis via additional rotation matrices $\hat{R}_i$ defined as $\frac{1}{2}\mu_B \hat{R}_i\mathbf{B}\cdot\hat{g}_i\boldsymbol{\sigma}_i=\frac{1}{2}E_{Z,i}{\sigma}^Q_{Z,i},i \in \{1, 2\}$. The resulting Hamiltonian is then given by
\begin{align}
    \mathcal{H}^Q = \frac{1}{2}E_{Z,1}{\sigma}^Q_{Z,1}
    + \frac{1}{2}E_{Z,2}{\sigma}^Q_{Z,2}
    + \frac{1}{4}\boldsymbol{\sigma}^Q_1\cdot \mathcal{J}^Q\boldsymbol{\sigma}^Q_2,
    \label{eq:QubitframeHam}
\end{align}
where $\mathcal{J}^Q=J_0\hat{R}_1\hat{R}_{\mathbf{n}_{\text{so}}}(-2\phi_{\text{so}})\hat{R}^T_2$. We refer to this frame as the qubit frame~\cite{geyerAnisotropicExchangeInteraction2024}. In this frame, the Zeeman terms become diagonal. As a result, the Zeeman Hamiltonian takes the same form as commonly used for describing electron spin qubits, making this frame convenient for quantum-gate design, while the exchange interaction $\mathcal{J}^Q$ remains anisotropic. In the following, we show how this anisotropic exchange interaction can be exploited to implement two-qubit gates.

\subsection{Quantum gate and noise model}
\label{Section: Quantum gate and noise model}
Staring from the qubit-frame Hamiltonian Eq.~\ref{eq:QubitframeHam}, we transform into the rotating frame with respect to the Zeeman interaction and implementing the rotating-wave approximation (see Appendix~\ref{Appendix B}), the resulting Hamiltonian can be approximated as
\begin{equation}
    \mathcal{H}=\frac{1}{4}\begin{pmatrix}
    J_{zz}^{Q} & 0 & 0 & 0 \\
    0 & 2\Delta E_z - J_{zz}^{Q} & {J_{\perp}} & 0 \\
    0 & {J_{\perp}^{*}} & -2\Delta E_z - J_{zz}^{Q} & 0 \\
    0 & 0 & 0 & J_{zz}^{Q} 
    \label{eq:RWA Hamiltonian}
    \end{pmatrix},
\end{equation}
where $\Delta E_Z=E_{Z,1}-E_{Z,2}$ and $J_{\perp}=J_{xx}^{Q}+J_{yy}^{Q}+i(J_{xy}^{Q}-J_{yx}^{Q})$. $\Delta E_Z$ can only be ignored under the approximation $|J_{\perp}| \ll |J_{zz}^Q|,|\Delta E_Z| $. When this condition is satisfied, the impact of $\Delta E_Z$ can be compensated by applying single-qubit phase gates. The resulting time evolution operator generated by $\mathcal{H}$ can be reduced to $e^{-i J_{zz}^Q \sigma_z \sigma_z t/4}$ (for simplicity we use $\sigma$ to represent $\sigma^Q$ in the following). A target ZZ phase rotation $e^{-i\Theta\,\sigma_{z}\sigma_{z}}$ can then be generated by the corresponding time evolution with $t=4\Theta/J_{zz}^{Q}$. Alternatively, a CNOT gate can also be realized by applying a drive Hamiltonian~\cite{geyerAnisotropicExchangeInteraction2024,russHighfidelityQuantumGates2018,hendrickxFourqubitGermaniumQuantum2021}.

In the discussion above, we have assumed ideal conditions of no decoherence and a perfectly calibrated gate, where the $g$-tensors and the exchange coupling $\mathcal{J}$ remain fixed at their calibrated values throughout the execution of the two-qubit gate. In practice, however, experimental limitations and low frequency noise lead to parameter miscalibration and fluctuations, resulting in imperfect gate operations. In this work, we focus on the impact of charge noise on the exchange coupling, modeling it as a $\mathcal{J}$-tensor fractional error $J_0 \rightarrow J_0(1+\varepsilon)$, where $\varepsilon$ denotes a small fractional deviation of the coupling strength. We assume that this fractional error remains constant during a single gate operation but can vary between different runs, which corresponds to the quasi-static approximation. This assumption is well justified, since the duration of a single gate operation is much shorter than the characteristic timescale over which the noise varies~\cite{dialChargeNoiseSpectroscopy2013,martinsNoiseSuppressionUsing2016,burkardSemiconductorSpinQubits2023,paladinoNoiseImplicationsSolidstate2014,paquelet2023reducing,lodariLowPercolationDensity2021,wangOperatingSemiconductorQuantum2024a,stehouwer2025exploiting,hendrickxSweetspotOperationGermanium2024,piotSingleHoleSpin2022} as the dominant contribution arises from low-frequency components of the noise. Furthermore, charge traps at the semiconductor-oxide interface induce fluctuations in the number of carriers, ultimately resulting in voltage noise~\cite{martinezVariabilityElectronHole2022,martinezVariabilityMitigationEpitaxialheterostructurebased2024,martinez2025variability} which directly manifest as fluctuations in the applied electrostatic gate voltages. Such voltage fluctuations lead to variations in the inter-dot barrier height and impose a random bias potential between the two dots, which will also perturb the exchange coupling~\cite{burkardCoupledQuantumDots1999,huChargeFluctuationInducedDephasingExchangeCoupled2006}. 

To quantitatively assess the impact of these effects, we benchmark the gate performance using the average gate fidelity. Let $U$ denote the target two-qubit gate unitary and $V$ the actual time-evolution operator generated by the system dynamics. A simple equation for calculating the average gate fidelity is given by~\cite{white2007measuring}
\begin{align}
    F_{\text{avg}}=\frac{|\text{Tr}(U^\dagger V)|^2+d}{d(d+1)},
    \label{Average gate fidelity}
\end{align}
where $d=4$ denotes the dimension of the Hilbert space for a two-qubit system. A value of $F_{\text{avg}} = 1$ corresponds to an ideal implementation of the target gate, with performance improving as $F_{\text{avg}}$ approaches 1.

\section{Composite Pulse Scheme}
\label{Section: Composite poulse scheme}
To suppress low-frequency noise, we employ dedicated composite pulse sequences, where a single control pulse is substituted by a carefully designed sequence of pulses engineered to compensate gate errors arising from misscalibrations caused by experimental limitations or low-frequency electric fluctuations. As discussed in Sec.~\ref{Section: Quantum gate and noise model}, the dominant noise contributions arise from low-frequency spectral components, which can be efficiently suppressed using composite pulse schemes. In this work, we adopt a sequence known as SCROFULOUS~\cite{cumminsTacklingSystematicErrors2003}. For single-qubit gates, this sequence effectively suppresses the impact of rotation-angle deviation errors (also known as pulse-length errors) on gate performance. Fig.~\ref{fig:figure_1}(a) compares the gate fidelity of a single-pulse X gate and a SCROFULOUS composite-pulse X gate in the presence of pulse-length errors. Fig.~\ref{fig:figure_1}(b) illustrates on the Bloch sphere how a SCROFULOUS composite-pulse sequence compensates systematic rotation errors and yields a final state much closer to the intended target than a single-pulse gate. The two-qubit version of SCROFULOUS can be derived from the single-qubit sequence through an operator mapping between single- and two-qubit rotations~\cite{ichikawa2013minimal}. For two-qubit entangling gates implemented via the exchange interaction, this composite sequence efficiently compensates for systematic deviations in the exchange coupling amplitude.

The unitary operator sequence of a two-qubit version SCROFULOUS is~\cite{ichikawa2013minimal}
\begin{align}
    U= e^{-i\zeta \sigma_z \sigma_z} e^{i \frac{\theta}{2} I\sigma_x} e^{-i \frac{\pi}{2} \sigma_z \sigma_z} e^{-i \frac{\theta}{2} I\sigma_x} e^{-i\zeta \sigma_z \sigma_z}.
    \label{eq:SCROFULOUS}
\end{align}
By choosing $\zeta=-\frac{\pi}{4}\sec\theta, \sec\theta\approx-1.28$, this operator sequence is equivalent to $e^{-I \eta I \sigma_x} e^{-I \frac{\pi}{4} \sigma_z \sigma_z} e^{I \eta I \sigma_x}$ with $\tan{\eta}=\tan{\theta}\sec\left({\frac{\pi}{2}}\sec{\theta}\right)$~\cite{gungorduPulseSequenceDesigned2018}. This operation is thus locally equivalent to a ZZ-type entangling gate (CZ gate), up to two single-qubit rotations around the $x$-axis. We consider the case in which both the first and the last ZZ operators, $e^{-i\zeta \sigma_z \sigma_z}$, as well as the middle ZZ operator, $e^{-i \frac{\pi}{2} \sigma_z \sigma_z}$, are generated by the exchange interaction. Fluctuations of the exchange coupling originating from low-frequency charge noise lead to deviations in the accumulated ZZ phase during the gate operation. Owing to the appearance of the interleaved single-qubit X rotations in the sequence, these phase deviations are compensated. Consequently, the overall $e^{-i\frac{\pi}{4}\sigma_z \sigma_z}$ operator implemented via the SCROFULOUS sequence is robust against fractional $\mathcal{J}$-tensor amplitude errors.

\section{Noise-protected CZ Gate}
\label{Section: Noise-protected CZ gate}

\begin{figure}[t]
    \centering
    \includegraphics[width=0.4\textwidth]{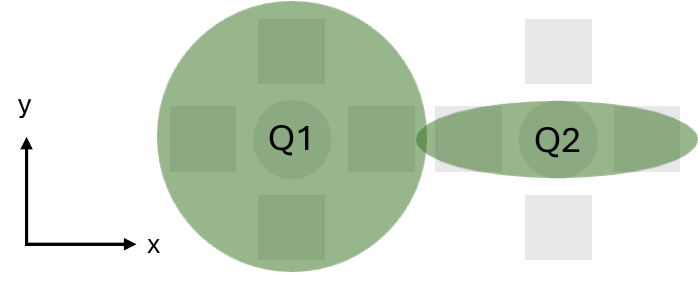}
    \caption{Schematic of the device architecture used in the operating scheme. Qubit 1 is tuned by pulsing the central gate voltage such that $g_{xx} = -g_{yy}$, while qubit 2 is squeezed along the $y$-direction resulting in $g_{yy} = 0$. The green regions represent the spatial extension of the two qubits’ wavefunctions, whose overlap gives rise to the exchange interaction.}
    \label{fig:figure_2}
\end{figure}

We now show how the SCROFULOUS composite pulse can be implemented using Ge hole spin qubits without additional single-qubit gates. As shown in sequence Eq.~\ref{eq:SCROFULOUS}, the first and last operators are identical, both taking the form $e^{-i\zeta\,\sigma_z\sigma_z}$, and can therefore be implemented using the method described in Sec.~\ref{Section: Quantum gate and noise model}. Instead of implementing separate single-qubit operations, the middle three operators: $e^{i \frac{\theta}{2} I\sigma_x} e^{-i \frac{\pi}{2} \sigma_z \sigma_z} e^{-i \frac{\theta}{2} I\sigma_x}$ can be obtained by a single evolution process (see Appendix~\ref{Appendix D}) if we suddenly change the Hamiltonian $\mathcal{H}^Q$ to
\begin{align}
    \mathcal{H}^{Q'}=
    &\frac{1}{2} E_{Z,1} \sigma_{Z,1} + \frac{1}{2}\mu_B 
    \hat{R}_2 \mathbf{B} \cdot \hat{g}_2  
    \left( R_x(\theta)\boldsymbol{\sigma}_2 \right) \notag \\
    & + \frac{1}{4}\,
    \boldsymbol{\sigma}_1 \cdot \mathcal{J}^Q 
    \left( R_x(\theta)\boldsymbol{\sigma}_2 \right).
\end{align}
This is equivalent to changing $\hat{g}_2 \rightarrow \hat{g}_2'$ such that
$ \hat{R}_2 \mathbf{B} \cdot \hat{g}_2 \left( R_x(\theta)\boldsymbol{\sigma}_2 \right) =  \hat{R}_2 \mathbf{B} \cdot \hat{g}_2' \boldsymbol{\sigma}_2$. Such a transformation of the g-tensor can be enabled by the electric tunability of the g-tensor.

\begin{figure}[t]
    \centering
    \includegraphics[width=0.45\textwidth]{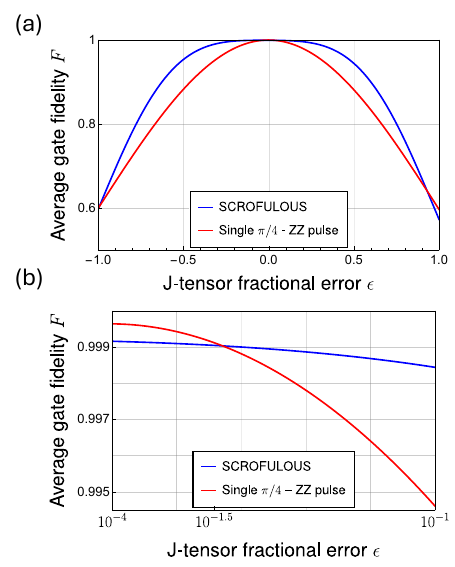}
    \caption{(a) Average gate fidelity comparison between the SCROFULOUS gate and a single-pulse $e^{-i\frac{\pi}{4}\sigma_z \sigma_z}$ gate realized by two Ge hole spin qubits in the presence of $\mathcal{J}$-tensor fractional error $\epsilon$. (b) Logarithmic scale magnification of the small-error region displayed in panel (a). The single-pulse ZZ$_{\pi/4}$ gate slightly outperforms the SCROFULOUS gate for extremely small $|\epsilon|$, but its fidelity decreases rapidly as $|\epsilon|$ increases.}
    \label{fig:figure_3}
\end{figure}

Such large sudden changes of the electrical g-tensor can for example be realized using hopping spins~\cite{wangOperatingSemiconductorQuantum2024a} or Gapless Single-spin (GS2) qubits~\cite{rimbach2025gapless}. Here, we focus on the latter architecture, which can be realized using planar Ge quantum dots. Fig.~\ref{fig:figure_2} shows the schematic of our device. Suppose we have two adjacent qubits, each controlled by a central gate and four side gates. Under the assumption of negligible strain, the g-tensor components are given by Eqs.~\ref{eq:g_xx}\,--\,\ref{eq:g_yz}. Assuming further a separable confinement, the g-tensor simplifies to $\mathrm{diag}(g_{xx},g_{yy},g_{zz})$. The gapless regime $g_{xx}=g_{yy}=0$ can be electrically tuned via $\langle p^2_x \rangle = \langle p^2_y \rangle = \frac{m_0 \Delta_{\mathrm{HL}} q}{2(\lambda-\lambda')}$. If we further set an in-plane magnetic field ($B_z=0$), the entire Zeeman Hamiltonian vanishes which leaves the qubit without energy splitting and become gapless. In the vicinity of this gapless regime, the g-tensor is highly tunable, and the required g-tensor transformation can be achieved with suitable adjustments of the gate voltages. Nevertheless, because the rotating wave approximation required to suppress spin-non-conserving terms relies on a finite Zeeman splitting, the Zeeman term cannot be too small. This prevents us from operating exactly at the gapless point. This condition can be maintained for example by electrically compressing the g-tensor along the x or y axis i.e., tuning $g_{yy}=0$ or $g_{xx}=0$.

Owing to the predominantly cubic SOI in planar hole systems~\cite{marcellina2017spin}, the spin-orbit angle is expected to be small. Therefore, we also assume a small spin-orbit angle in our model and approximate the $\mathcal{J}$-tensor in the lab frame by the identity matrix. We initialize the system such that qubit 1 operates in the fully gapless regime, while qubit 2 is compressed such that $g_{yy}=0$ as depicted in Fig.~\ref{fig:figure_2}. Numerically, we can identify a feasible set of parameters. These include the azimuthal angle $\phi$ of the in-plane magnetic field and the central gate voltage applied to qubit 1 before the gate operation, which ensures $|J_{zz}^{Q}|/|J_{\perp}|\gg 1$. The parameter set also includes the central-gate voltage applied to qubit 2 during the gate operation, which enables the required g-tensor transfer. Detailed data can be found in Appendix~\ref{Appendix C}. Fig.~\ref{fig:figure_3}(a) and (b) calculate and plot the average gate fidelity of a SCROFULOUS gate and the same $e^{-i\frac{\pi}{4}\sigma_z \sigma_z}$ gate realized by a single pulse in the presence of $\mathcal{J}$-tensor fractional error $\epsilon$. The gate fidelity is further numerically optimized by applying additional phase corrections at the beginning and the end of the pulse sequence. When $|\epsilon|$ starts to increase, the SCROFULOUS gate maintains a higher fidelity while the fidelity of the single-pulse gate drops rapidly. The intersection point between the two gate-fidelity curves at small $|\epsilon|$ shown in Fig.~\ref{fig:figure_3}(b) will shift to the left as the ratio $|J_{zz}^Q|/|J_{\perp}|$ increases. In the following, we investigate two additional sources of error affecting the gate performance, namely voltage noise and adiabatic errors.

\subsection{Voltage Noise}
\label{Section: Voltage Noise}
\begin{figure*}[t]
    \centering
    \includegraphics[width=1.01\textwidth]{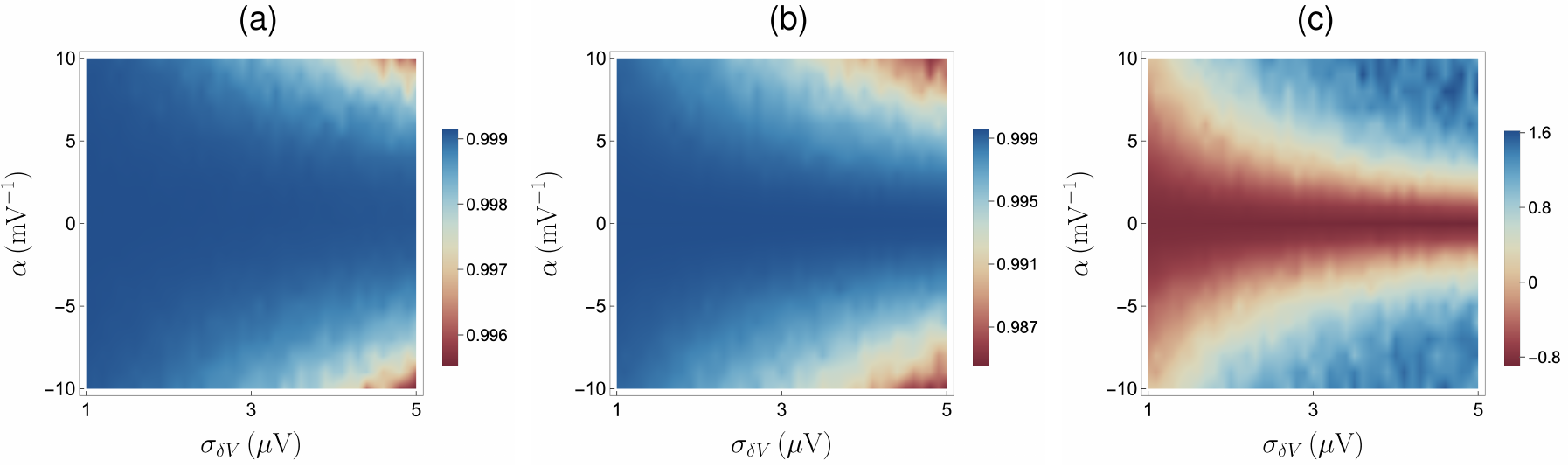}
    \caption{Comparison of gate fidelities in the presence of voltage noise affecting both the g-tensors and the exchange coupling. The fidelities are calculated using a Monte Carlo simulation, based on the details provided in Appendix~\ref{Appendix C}. (a) Gate-fidelity density plot for the SCROFULOUS gate. (b) Gate-fidelity density plot for the single-pulse ZZ gate. (c) Difference in logarithmic gate infidelity between the two protocols, defined as $\ln(1 - F_{\text{Single}}) - \ln(1 - F_{\text{SCROFULOUS}})$ and evaluated at identical $\sigma_{\delta V}$ and $\alpha$.  Positive values correspond to parameter regimes in which the SCROFULOUS gate outperforms the single-pulse ZZ gate.}
    \label{fig:figure_4}
\end{figure*}

As mentioned in Sec.~\ref{Section: Quantum gate and noise model}, charge traps at the semiconductor-oxide interface induce fluctuations in the applied electrostatic gate voltages. We model its effect on the exchange interaction by assuming an exponential dependence of the coupling constant on the voltage fluctuation~\cite{rimbach-russSimpleFrameworkSystematic2023,jirovec2025mitigation} and perform a Taylor expansion
\begin{align}
  J(\delta V_1, \delta V_2, \delta V_3)
    &= J_0 \,
       e^{2\alpha(\delta V_1+\delta V_2+\delta V_3)}\nonumber \\
    &\approx J_0
       \left(
         1+2\alpha(\delta V_1+\delta V_2+\delta V_3)
       \right).
       \label{eq:voltage_noise_exchange}
\end{align}
Here $\alpha$ is a leverarm, $\delta V_1$ and $\delta V_2$ denote the amplitudes of voltage noise applied to the central gates of qubit 1 and qubit 2 respectively, while $\delta V_3$ is introduced to effectively capture additional voltage noises that collectively affect the exchange coupling. We assume that $\delta V_1$, $\delta V_2$, and $\delta V_3$ are three independent random variables following Gaussian distributions with zero mean and the same standard deviation $\sigma_{\delta V}$. A Monte Carlo simulation is performed over these variables. For each individual noise realization, we model the effect of voltage noise on the central gates by making the g-tensors of the two qubits depend on the local voltage fluctuations, $\hat g_i = \hat g_i(\delta V_i), i = 1,2$. The exchange coupling is then evaluated according to Eq.~\ref{eq:voltage_noise_exchange} for a fixed $\alpha$ and $\sigma_{\delta V}$, and the resulting gate fidelity is calculated. The mean fidelity is then obtained by averaging over $300$ independent noise realizations. We further sweep both $\alpha$ and $\sigma_{\delta V}$ to systematically characterize their impact on the gate performance. By comparing the gate fidelity of our SCROFULOUS gate and a single-pulse ZZ gate as a function of noise strength and leverarm, we confirm the quantitative differences. Fig.~\ref{fig:figure_4}(a) and (b) show the fidelity density plots for the SCROFULOUS and single-pulse ZZ gates, respectively. The direct comparison in Fig.~\ref{fig:figure_4}(c) shows that when the voltage noise becomes sufficiently strong to induce substantial fluctuations in the exchange coupling, the SCROFULOUS gate protocol outperforms the single-pulse ZZ gate. This result further demonstrates its enhanced robustness and noise-protection capability.
\begin{figure*}[t]
    \centering
    \includegraphics[width=1\textwidth]{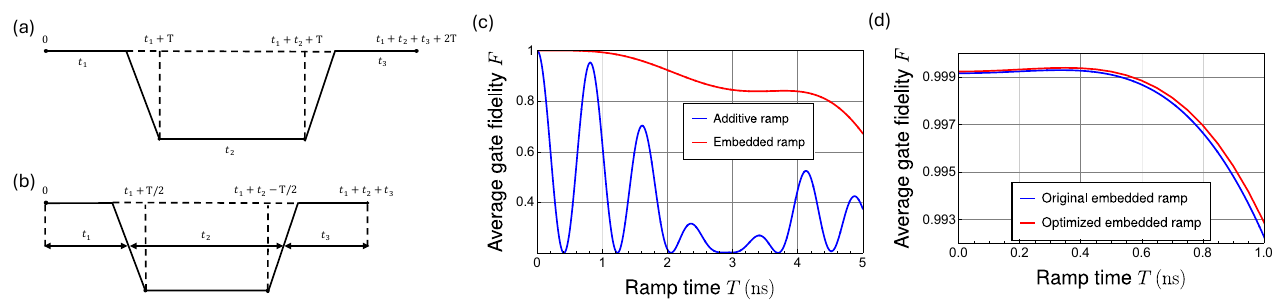}
    \caption{(a) Voltage pulse with two additive ramp segments of duration $T$. The total duration of the pulse sequence is $t_1 + t_2 + t_3 + 2T$. (b) Voltage pulse with embedded ramp time $T$, where the two ramps are incorporated into the three SCROFULOUS gate segments $t_1$, $t_2$, and $t_3$. The total duration remains $t_1 + t_2 + t_3$ and the pulse takes a symmetric shape. The durations $t_1$, $t_2$, and $t_3$ are specified in Appendix~\ref{Appendix C}. (c) Average gate fidelity as a function of the ramp time $T$ for the two ramping protocols shown in (a) and (b): pulse with additive ramps (blue) and embedded ramps (red). The embedded-ramp pulse exhibits a significant enhancement in gate fidelity, without showing rapid oscillations. (d) Average gate fidelity comparison between the original embedded-ramp pulse and the optimized pulse.}
    \label{fig:figure_5}
\end{figure*}

\begin{figure}[t]
    \centering
    \includegraphics[width=0.45\textwidth]{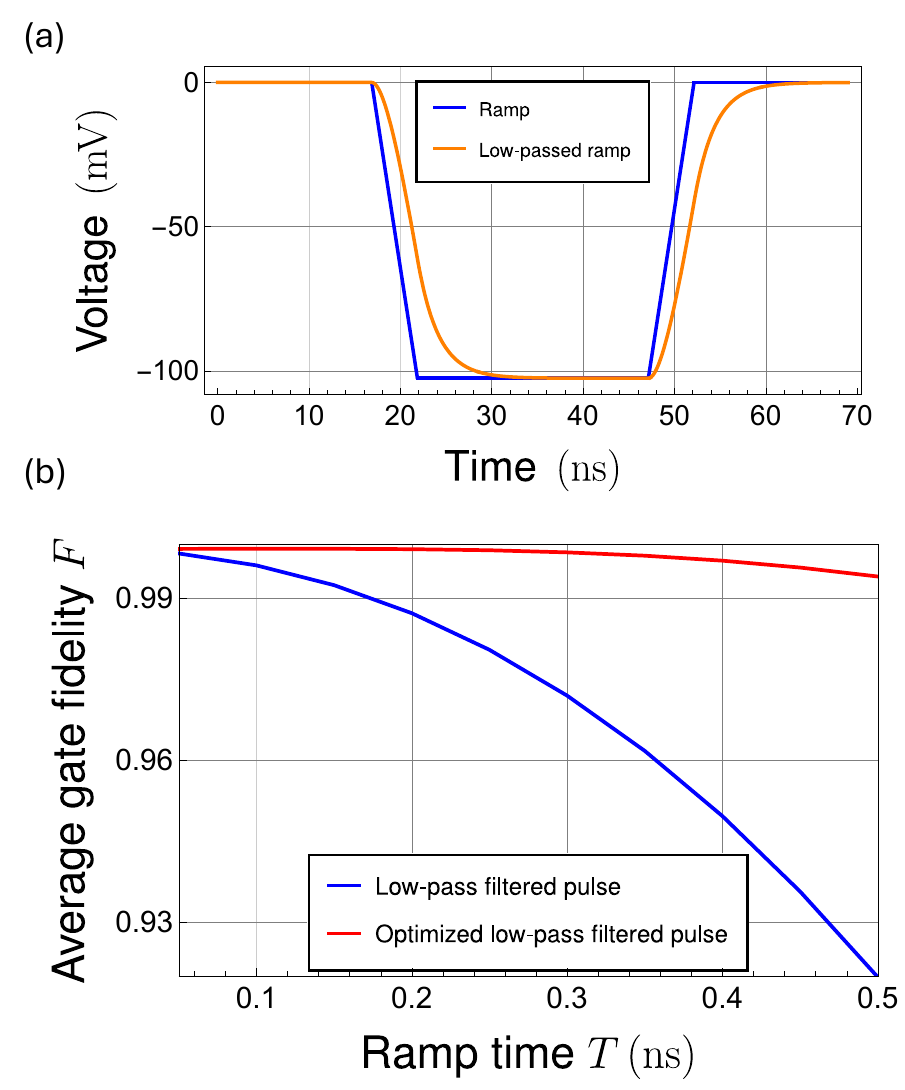}
    \caption{(a) Example of a embedded-ramp voltage pulse (blue) with a ramp time $T=5\,\mathrm{ns}$, together with the corresponding pulse after passing through a low-pass filter (orange). (b) Average gate fidelity comparison between the original low-pass filtered pulse and the optimized pulse.}
    \label{fig:figure_6}
\end{figure}

\subsection{Adiabatic Error }
\label{Section: Adiabatic Error}
So far our analysis was assumed that the voltage on central gates of qubit 2 can be instantaneously tuned to our preset value. In practice, however, implementing such ultrafast voltage transitions imposes stringent demands on control electronics. The voltage pulses in real experiments are expected to exhibit a finite transition time $T$, rather than ideal instantaneous square pulses. We first consider simplified linear ramp voltage pulse. As shown in Fig.~\ref{fig:figure_5}(a) and (c), when the final pulse shape explicitly includes both rising and falling ramps with duration $T$, such that the total pulse duration becomes $t_1 + t_2 + t_3 + 2T$, the average gate fidelity exhibits rapid oscillations as a function of ramp time. This behavior arises from the time evolution of the Zeeman terms, which induces high-frequency phase oscillations that strongly degrade the gate performance. To mitigate the phase oscillations, we instead incorporate the ramp time $T$ into the three SCROFULOUS gate segments $t_1$, $t_2$, and $t_3$, while maintaining a symmetric pulse shape, as illustrated in Fig.~\ref{fig:figure_5}(b). In this embedded-ramp scheme, the total gate duration remains $t_1 + t_2 + t_3$ while the pulse symmetry enables partial phase compensation. As evidenced by the fidelity plots, this embedded-ramp pulse yields a significant improvement over the additive-ramp scheme, with the gate fidelity remaining high over a finite range of ramp times. From Fig.~\ref{fig:figure_5}(d), we observe that the fidelity of the embedded-ramp pulse remains above $0.99$ even when the ramp time $T$ reaches $1~\mathrm{ns}$. Since the fidelity degradation primarily originates from phase errors, the gate performance can be further improved by numerically adjusting the durations of the three individual pulses in the SCROFULOUS sequence to compensate for these phase errors. Fig.~\ref{fig:figure_5}(d) shows the optimized gate fidelity. Remarkably, only slight improvements are possible with the optimized ramps compared to our simple scheme.

In realistic experimental settings, the finite bandwidth of the control electronics results in an effective low-pass filtering of the applied voltage pulse. In this work, we model this effect using a RC low-pass filter, as seen Fig.~\ref{fig:figure_6} and discussed in Appendix~\ref{Appendix E}. After filtering, the pulse becomes smoother; however, the associated phase delay renders the pulse shape asymmetric, as illustrated in Fig.~\ref{fig:figure_6}(a). Fig.~\ref{fig:figure_6}(b) shows that the gate fidelity in this case decreases slightly faster than the linear case. We note that the low-pass-filtered waveform still depends on the original gate durations $t_1$, $t_2$, and $t_3$. Consequently, we again numerically adjust the durations of the three segments in the sequence to compensate for the phase errors and optimize the gate fidelity. As shown in Fig.~\ref{fig:figure_6}(b), this optimization leads again to a significant improvement in performance, with the gate fidelity remaining above $0.99$ for a ramp time of $T = 0.5~\mathrm{ns}$. Using the relation $\tau = T/\ln 9$ provided in Appendix~\ref{Appendix E}, we obtain a corresponding cutoff frequency of approximately $700~\mathrm{MHz}$ for $T = 0.5~\mathrm{ns}$. Appendix~\ref{Appendix F} presents the numerically obtained corrections $\Delta t_1$, $\Delta t_2$, and $\Delta t_3$ for the embedded-ramp pulse and the low-pass–filtered pulse. 

\subsection{Spectral Noise}
\label{Section: Spectral Noise}
\begin{figure}[t]
    \centering
    \includegraphics[width=0.45\textwidth]{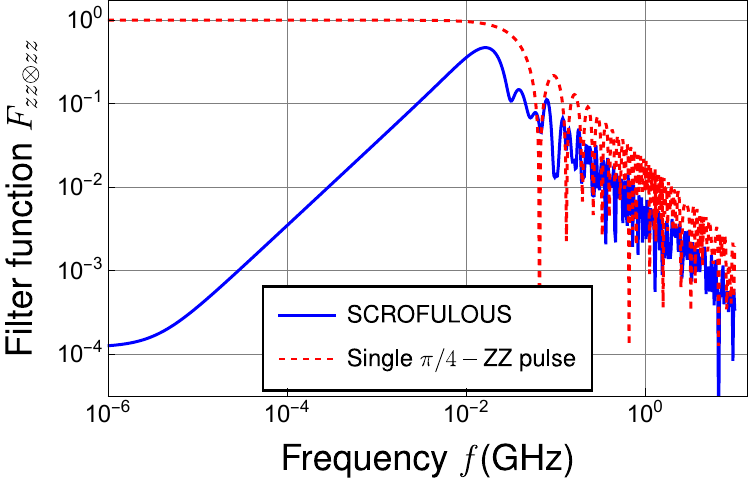}
    \caption{Filter function $F_{zz\otimes zz}(f)$ plotted on a logarithmic scale for the SCROFULOUS (blue solid) and a single $e^{-i\frac{\pi}{4}\sigma_z \sigma_z}$ pulse (red dashed). In the low-frequency regime, the SCROFULOUS filter function is strongly suppressed compared to the single-pulse case, indicating a substantially reduced overlap with low-frequency noise.}
    \label{fig:figure_7}
\end{figure}

So far, we have assumed that the $\mathcal{J}$-tensor fractional error $\varepsilon$ remains constant during a gate operation but varies between realizations which corresponds to the quasi-static error. We further investigate the robustness of our SCROFULOUS gate against arbitrary spectral noise using the filter function formalism~\cite{greenArbitraryQuantumControl2013}. This formalism provides a qualitative approach for characterizing the noise susceptibility of quantum control protocols such as composite pulses and dynamical decoupling sequences. The noise-suppression capability (coherence) of a given control protocol is determined by the overlap between its filter function and the noise's power spectral density: the smaller the overlap, the more effectively the protocol mitigates the noise~\cite{greenArbitraryQuantumControl2013,hansenAccessingFullCapabilities2023}. Fig.~\ref{fig:figure_7} shows the filter function for SCROFULOUS and a single $e^{-i\frac{\pi}{4}\sigma_z \sigma_z}$ pulse, assuming an identical noise channel proportional to $\sigma_z \sigma_z$. Both filter functions are normalized with respect to their own total gate durations. We can clearly observe that our SCROFULOUS gate suppresses the impact of low-frequency noise, such as the omnipresent 1/f noise in semiconductors, as its filter function exhibits minimal overlap with the slowly varying components of the noise spectrum. 

\section{Conclusion and Outlook}
\label{Section: Conclusion and Outlook}
We have presented a noise-protected two-qubit gate protocol for planar Ge hole spin qubits without the need of additional sinle-qubit operations, exploiting the electrical tunability of g-tensor and the anisotropy of exchange interaction. By implementing the SCROFULOUS composite pulse sequence, our protocol realizes a robust universal CZ gate. Numerical simulations confirm that the composite-pulse gate effectively suppresses quasi-static fluctuations of the exchange coupling constant caused by charge noise and outperforms a conventional single-pulse ZZ gate under the same noise conditions. We further showed the robustness against realistic low-frequency noise via the filter function formalism. Additional error sources, such as voltage noise and adiabatic errors, can also be suppressed using simple pulse time compensation techniques. 

Among composite-pulse gates based on Ising-type interactions, SCROFULOUS provides a minimal entangling construction with three segments and first-order error cancellation~\cite{ichikawa2013minimal}. Higher-order suppression can be achieved with more sophisticated schemes such as the broadband-1 (BB1)~\cite{wimperis1994broadband,tomita2010multi}. Concatenated composite pulses that simultaneously suppress different types of systematic errors are also possible~\cite{bandoConcatenatedCompositePulses2013}. Meanwhile, this noise-protection strategy is not restricted to the CZ gate studied here, but has broader applicability. By engineering the exchange-coupling tensor and tailoring appropriate voltage pulse sequences, similar composite-pulse gate protocols can be designed to realize other high-fidelity entangling operations, such as iSWAP-type gates. Our results highlight a viable pathway toward scalable and fault-tolerant quantum processors based on electrically controlled Ge hole spin qubits.

\section{Acknowledgement}
We thank all members of the Bosco, Rimbach-Russ, Scappucci, Veldhorst, and Vandersypen group for valuable feedback. We also thank P. Harvey-Collard and L. Massai for insightful discussions and comments. We also thank M.T.P. Nguyen for critical feedback to the manuscript. M.R.-R. acknowledges support from the Dutch Research Council (NWO) under Award Number Vidi TTW 22204. This research was supported by the H2024 QLSI2 project and the Army Research Office under Award Number: W911NF-23-1-0110. 
The views and conclusions contained in this document are those of the authors and should not be interpreted as representing the official policies, either expressed or implied, of the Army Research Office or the U.S. Government. The U.S. Government is authorized to reproduce and distribute reprints for Government purposes notwithstanding any copyright notation herein.

\section{Data Availability}
The dataset that supports the findings of this article are openly available at~\footnote{Public repository found at \url{https://doi.org/10.5281/zenodo.18956880}.}.

\makeatletter
\renewcommand\appendixname{APPENDIX}
\makeatother
\appendix

\section{EXCHANGE INTERACTION IN THE LAB AND SPIN-ORBIT FRAME}
\label{Appendix A}
In the second quantization formalism, the Pauli operators 
acting on the orbital and spin degrees of freedom can be 
written in terms of their matrix elements as follows. 
For the orbital subspace ($i,j \in \{1,2\}$), the corresponding operators are given by
\begin{subequations}\label{eq:tau_def}
\begin{align}
  ( \tau_x )_{ij} 
  &= \delta_{i1}\delta_{j2} + \delta_{i2}\delta_{j1}, \\
  ( \tau_y )_{ij} 
  &= -i\,\delta_{i1}\delta_{j2} + i\,\delta_{i2}\delta_{j1}, \\
  ( \tau_z )_{ij} 
  &= \delta_{i1}\delta_{j1} - \delta_{i2}\delta_{j2}.
\end{align}
\end{subequations}
Accordingly, the projection operators are given by
\begin{subequations}\label{eq:projectors}
\begin{align}
  \left( \frac{1 + \tau_z}{2} \right)_{ij}
  &= \delta_{i1}\delta_{j1}, \\
  \left( \frac{1 - \tau_z}{2} \right)_{ij}
  &= \delta_{i2}\delta_{j2}.
\end{align}
\end{subequations}
For the spin subspace ($s,s' \in \{\uparrow,\downarrow\}$), 
the Pauli vector $\boldsymbol{\sigma}$ has components
\begin{subequations}\label{eq:sigma_def}
\begin{align}
  ( \sigma_x )_{ss'} 
  &= \delta_{s\uparrow}\delta_{s'\downarrow}
  + \delta_{s\downarrow}\delta_{s'\uparrow}, \\
  ( \sigma_y )_{ss'} 
  &= -i\,\delta_{s\uparrow}\delta_{s'\downarrow}
  + i\,\delta_{s\downarrow}\delta_{s'\uparrow}, \\
  ( \sigma_z )_{ss'} 
  &= \delta_{s\uparrow}\delta_{s'\uparrow}
  - \delta_{s\downarrow}\delta_{s'\downarrow}.
\end{align}
\end{subequations}

To find the effective low-energy dynamics of the DQD Hamiltonian Eq.~\ref{eq:DQD Hamiltonian}, we first introduce here the spin-orbit frame, in which the spin-flip tunneling term is removed by local spin rotations~\cite{geyerAnisotropicExchangeInteraction2024}. We then discuss the conditions under which the system can be restricted to the $(1,1)$ charge stability regime. Afterwards, we introduce localized spin operators to rewrite our effective Hamiltonian into a Heisenberg spin Hamiltonian.

The spin-orbit frame is implemented via the unitary transformation $U_{\text{so}}=\exp\left(-i\phi_{\text{so}}\tau_z\mathbf{n}_{\text{so}}\cdot\boldsymbol{\sigma}/2\right)$, which induces opposite rotations of the spin-quantization axes at the two sites, such that the two spins acquire opposite phases. Under this transformation, the DQD Hamiltonian becomes: $\mathcal{H}^{\text{so}}_{\text{DQD}}=U_{\text{so}}^{\dagger}\mathcal{H}_{\text{DQD}}U_{\text{so}}$. The on-site interaction term $U n_{i\uparrow} n_{i\downarrow}$ remains unchanged, while the remaining part $\bar{\mathcal{H}}$ is transformed into
\begin{align}
    \bar{\mathcal{H}}^{\text{so}} = \frac{1}{2}\mu_B \mathbf{B}\cdot\left[
        \frac{1 + \tau_z}{2}\,\hat{g}^{\text{so}}_1\boldsymbol{\sigma}
        + \frac{1 - \tau_z}{2}\,\hat{g}^{\text{so}}_2\boldsymbol{\sigma}
        \right] + \frac{\epsilon_{\text{eff}}}{2}\tau_z + t\tau_x.
    \label{ham:so_dqd}
\end{align}
During the derivation, we make use of the SU(2)-SO(3) isomorphism
\begin{align}
    U^\dagger \boldsymbol{\sigma} U
    = \hat{R}\,\boldsymbol{\sigma},
\end{align}
where $\hat{R}$ is a 3D rotation matrix acting on the 3 vector components. As a result, $\hat{g}^{\text{so}}_1\boldsymbol{\sigma}=\hat{g}_1\hat{R}_{\mathbf{n}_{\text{so}}}(\phi_{\text{so}})\boldsymbol{\sigma}$ and $\hat{g}^{\text{so}}_2\boldsymbol{\sigma}=\hat{g}_2\hat{R}_{\mathbf{n}_{\text{so}}}(-\phi_{\text{so}})\boldsymbol{\sigma}$. Therefore, we can naturally define $\hat{g}^{\text{so}}_1=\hat{g}_1\hat{R}_{\mathbf{n}_{\text{so}}}(\phi_{\text{so}})$ and $\hat{g}^{\text{so}}_2=\hat{g}_2\hat{R}_{\mathbf{n}_{\text{so}}}(-\phi_{\text{so}})$.

The basis states $\big\{\,|S(0,2)\rangle,\; |S\rangle,\; |T_{\uparrow\uparrow}\rangle,\; |T_{\downarrow\downarrow}\rangle,\; |T_0\rangle\,\big\}$ are defined as (using the notation $\lvert n_{1\uparrow}, n_{1\downarrow}; n_{2\uparrow}, n_{2\downarrow}\rangle$)
\begin{subequations}\label{eq:two_electron_basis}
\begin{align}
  \lvert S(0,2)\rangle
  &= \lvert 0,0;\,1_{\uparrow},1_{\downarrow}\rangle, \\
  \lvert S\rangle
  &= \frac{1}{\sqrt{2}}
  \Bigl(
    \lvert 1_{\uparrow},0;\,0,1_{\downarrow}\rangle
    -
    \lvert 0,1_{\downarrow};\,1_{\uparrow},0\rangle
  \Bigr), \\
  \lvert T_0\rangle
  &= \frac{1}{\sqrt{2}}
  \Bigl(
    \lvert 1_{\uparrow},0;\,0,1_{\downarrow}\rangle
    +
    \lvert 0,1_{\downarrow};\,1_{\uparrow},0\rangle
  \Bigr), \\
  \lvert T_{\uparrow\uparrow}\rangle
  &= \lvert 1_{\uparrow},0;\,1_{\uparrow},0\rangle, \\
  \lvert T_{\downarrow\downarrow}\rangle
  &= \lvert 0,1_{\downarrow};\,0,1_{\downarrow}\rangle.
\end{align}
\end{subequations}
Projecting the spin-orbit frame DQD Hamiltonian to this basis, we can write it into the following matrix form:
\begin{align}
    {
    \mathcal{H}_{5\times5}=
    \begin{pmatrix}
        U_0 - \epsilon
        & \sqrt{2}\,t
        & 0
        & 0
        & 0 \\
        \sqrt{2}\,t
        & 0
        & -\frac{\delta{b}_{+}}{\sqrt{2}}
        & \frac{\delta{b}_{-}}{\sqrt{2}}
        & \delta b_z \\
        0
        & -\frac{\delta{b}_{-}}{\sqrt{2}}
        & \bar{b}_z
        & 0
        & \frac{\bar{b}_{-}}{\sqrt{2}} \\
        0
        & \frac{\delta{b}_{+}}{\sqrt{2}}
        & 0
        & -\bar{b}_z
        & \frac{\bar{b}_{+}}{\sqrt{2}} \\
        0
        & \delta b_z
        & \frac{\bar{b}_{+}}{\sqrt{2}}
        & \frac{\bar{b}_{-}}{\sqrt{2}}
        & 0
    \end{pmatrix}}.
\end{align}
We define $\bar{b}_{\pm}=\bar{b}_x \pm i\bar{b}_y$ and $\delta b_{\pm}=\delta b_x \pm i\delta b_y$. Note here we work near the $S(0,2)-S$ anticrossing which allows us to omit the $S(2,0)$ state that has a much higher energy. The average and gradient Zeeman fields are introduced as
$\mathbf{\bar{b}} = (\mathbf{b}_1 + \mathbf{b}_2)/2$ and $\delta\mathbf{b} = (\mathbf{b}_1 - \mathbf{b}_2)/2$, with the local Zeeman field on site i defined by $\mathbf{b}_i = \mu_B\,\mathbf{B}\cdot\hat{g}^{\text{so}}_i$.

The upper-left $2\times2$ block of the matrix is spanned by the states $\big\{\,|S(0,2)\rangle,\; |S\rangle\,\big\}$. By transforming to another basis states set: $\big\{\,|S_{+}\rangle,\; |S_{-}\rangle\,\big\}$ which relates to $\big\{\,|S(0,2)\rangle,\; |S\rangle\,\big\}$ by
\begin{align}
    \begin{bmatrix}
    S_{+} \\
    S_{-}
    \end{bmatrix} =
    \begin{pmatrix}
    \cos\frac{\gamma}{2} & \sin\frac{\gamma}{2}\\
    -\sin\frac{\gamma}{2} & \cos\frac{\gamma}{2}
    \end{pmatrix}
    \begin{bmatrix}
    S(0,2) \\
    S
    \end{bmatrix},
\end{align}
we can diagonalize the matrix by choosing $\tan\gamma = 2\sqrt{2}t/(U_0 - \epsilon)$. Rewriting the $\mathcal{H}_{5\times5}$ in the basis states $\big\{\,|S_{+}\rangle,\; |S_{-}\rangle,\; |T_{\uparrow\uparrow}\rangle,\; |T_{\downarrow\downarrow}\rangle,\; |T_0\rangle\,\big\}$ and omitting the coupling between the $S_{+}$ and the triplet states because $\sin(\gamma/2)$ is very small, we can arrive at a $4\times4$ Hamiltonian in the basis $[\big\{\,|S_{-}\rangle,\; |T_{\uparrow\uparrow}\rangle,\; |T_{\downarrow\downarrow}\rangle,\; |T_0\rangle\,\big\}]$
\begin{align}
    {
    \begin{pmatrix}
        -J_0
        & -\frac{\delta{b}_{+}}{\sqrt{2}}\cos({\frac{\gamma}{2}})
        & \frac{\delta{b}_{-}}{\sqrt{2}}\cos({\frac{\gamma}{2}})
        & \delta b_z\cos({\frac{\gamma}{2}}) \\
        -\frac{\delta{b}_{-}}{\sqrt{2}}\cos({\frac{\gamma}{2}})
        & \bar{b}_z
        & 0
        & \frac{\bar{b}_{-}}{\sqrt{2}} \\
        \frac{\delta{b}_{+}}{\sqrt{2}}\cos({\frac{\gamma}{2}})
        & 0
        & -\bar{b}_z
        & \frac{\bar{b}_{+}}{\sqrt{2}} \\
        \delta b_z\cos({\frac{\gamma}{2}})
        & \frac{\bar{b}_{+}}{\sqrt{2}}
        & \frac{\bar{b}_{-}}{\sqrt{2}}
        & 0
    \end{pmatrix}},
\end{align}
where $J_0 = \sqrt{2}t \tan\left(\frac{\gamma}{2}\right)$. 

The $\sin(\gamma/2)$ is very small also indicates the Zeeman fields after renormalization are very close to the original ones~\cite{geyerAnisotropicExchangeInteraction2024}. Notice that $S_{-} \approx S$ in this week tunneling regime and the resulting $4 \times 4$ Hamiltonian falls entirely into the (1,1) regime, in the basis $\big\{\,|S\rangle,\; |T_{\uparrow\uparrow}\rangle,\; |T_{\downarrow\downarrow}\rangle,\; |T_0\rangle\,\big\}$, the $4 \times 4$ Hamiltonian can be rewritten as
\begin{align}
    \mathcal{H}^{\text{so}} = \frac{1}{2}\mu_B\mathbf{B}\cdot\hat{g}^{\text{so}}_1\boldsymbol{\sigma}^{\text{so}}_1 + \frac{1}{2}\mu_B\mathbf{B}\cdot\hat{g}^{\text{so}}_2\boldsymbol{\sigma}^{\text{so}}_2 + \frac{1}{4}J_0\boldsymbol{\sigma}^{\text{so}}_1\cdot\boldsymbol{\sigma}^{\text{so}}_2,
    \label{ham:so}
\end{align}
where the $\boldsymbol{\sigma}^{\text{so}}_{i},i\in\{1,2\}$ represent localized spin operators in this spin-orbit frame. The exchange Hamiltonian $\frac{1}{4}J_0\,\boldsymbol{\sigma}^{\text{so}}_1\!\cdot\!\boldsymbol{\sigma}^{\text{so}}_2$ is actually $\mathrm{diag}(-\tfrac{3}{4}J_0,\,\tfrac{1}{4}J_0,\,\tfrac{1}{4}J_0,\,\tfrac{1}{4}J_0)$
in the $\big\{\,|S\rangle,\; |T_{\uparrow\uparrow}\rangle,\; |T_{\downarrow\downarrow}\rangle,\; |T_0\rangle\,\big\}$ basis, corresponding to the singlet and triplet manifolds. Here we shift the energy reference by $-\tfrac{1}{4}J_0$, taking the triplet manifold as the zero-energy level, so that the Hamiltonian becomes
$\mathrm{diag}(-J_0,\,0,\,0,\,0)$. In the spin-orbit frame, the exchange interaction is isotropic, but the $g$-tensor in the Zeeman term does not correspond directly to experimentally measured values due to the applied basis transformation. Defining $\boldsymbol{\sigma}_1=\hat{R}_{\mathbf{n}_{\text{so}}}(-\phi_{\text{so}})\boldsymbol{\sigma}^{\text{so}}_1,\boldsymbol{\sigma}_2=\hat{R}_{\mathbf{n}_{\text{so}}}(\phi_{\text{so}})\boldsymbol{\sigma}^{\text{so}}_2$, we obtain the lab-frame Hamiltonian Eq.~\ref{eq: lab-frame Hamiltonian}.

\section{CZ GATE CALIBRATION}
\label{Appendix B}
Starting from $\mathcal{H}^Q$, consider the rotating frame with respect to the total Zeeman energy defined by the unitary operator 
\begin{align}
    U_{\text{rf}}(t) = \exp\left(-i\omega\left(\sigma_{Z,1}+{\sigma}_{Z,2}\right )t/2\right),
\end{align}
with $\omega=(E_{Z,1}+E_{Z,2})/2$. $\mathcal{H}^Q$ in this rotating frame will be transformed into
\begin{align}
    \mathcal{H}^{Q}_{\text{rf}} & = U^\dagger_{\text{rf}}(t) \mathcal{H}^Q U_{\text{rf}}(t) - i U^\dagger_{\text{rf}}(t) \left( \frac{d}{dt} U_{\text{rf}}(t) \right ) \nonumber \\
    & =\frac{1}{4}\left(E_{Z,1}-E_{Z,2}\right)\sigma_{Z,1} + \frac{1}{4}\left(E_{Z,2}-E_{Z,1}\right)\sigma_{Z,2} \nonumber \\
    & \quad + U^\dagger_{\text{rf}}(t) \left(\frac{1}{4}\boldsymbol{\sigma}_1\cdot \mathcal{J}^Q\boldsymbol{\sigma}_2 \right) U_{\text{rf}}(t).
\end{align}
We focus on the $U^\dagger_{\mathrm{rf}}(t)\left(\frac{1}{4}\boldsymbol{\sigma}_1\cdot\mathcal{J}^Q\boldsymbol{\sigma}_2\right)U_{\mathrm{rf}}(t)$ term in the matrix representation. We find that all off-diagonal elements, except those within the $\{|\!\!\uparrow\downarrow\rangle,|\!\!\downarrow\uparrow\rangle\}$ subspace, acquire time-dependent phase factors $e^{\pm i\omega t}$ or $e^{\pm 2 i\omega t}$. When $|J_{zz}^Q|/|J_{\perp}| \gg 1$ is satisfied and the rapidly oscillating off-diagonal elements are neglected, $\mathcal{H}^{Q}_{\text{rf}}$ reduces to
\begin{align}
    \mathcal{H}^{Q}_{\text{rf}} 
    =\frac{1}{4}\left(\Delta E_Z\sigma_{Z,1} - \Delta E_Z\sigma_{Z,2} + J_{zz}^Q \sigma_{Z,1} \sigma_{Z,2}\right).
\end{align}
with $\Delta E_Z=E_{Z,1}-E_{Z,2}$.
Since the resulting Hamiltonian contains only Pauli-Z operators, $J_{zz}^Q \sigma_{Z,1} \sigma_{Z,2}$ commutes with the two phase terms containing $\sigma_{Z,1}$ and $\sigma_{Z,2}$. Consequently, the first two unwanted phase terms can be straightforwardly canceled through phase compensation, as discussed below.

Assume $t_1=4\zeta/J_{zz}^Q$ is evolution time required to implement the first operator in the SCROFULOUS sequence: $e^{-i\zeta \sigma_z \sigma_z}$. The time evolution of the $\mathcal{H}^{Q}_{\text{rf}}$ is given by
\begin{align*}
    U^{Q}_{\text{rf}}(t_1)
    &=e^{-i \mathcal{H}^{Q}_{\text{rf}} t_1}\\
    &=e^{-i\Delta E_Z\left(\sigma_{Z,1} - \sigma_{Z,2}\right)t_1/4} e^{-iJ_{zz}^Q \sigma_{Z,1} \sigma_{Z,2}t_1/4}\\
    &=e^{-i\Delta E_Z\left(\sigma_{Z,1} - \sigma_{Z,2}\right)t_1/4} e^{-i\zeta \sigma_{Z,1} \sigma_{Z,2}}.
\end{align*}
The unwanted single-qubit phase accumulated during this evolution can therefore be compensated by applying the phase operation $\exp\left[i\Delta E_Z\left(\sigma_{Z,1}-\sigma_{Z,2}\right)t/4\right]$. If one further transforms from the rotating frame back to the original qubit frame, this can be achieved by an additional phase compensation $\exp\left[i\omega \left(\sigma_{Z,1}+{\sigma}_{Z,2}\right )t_1/2\right]$. Since the initial and final rotation operators in the SCROFULOUS sequence are identical, the required phase compensations can be applied before the sequence begins and after the operation ends.

\section{NUMERICAL SIMULATION DETAILS}
\label{Appendix C}
Before presenting details of the gate implementation, we first specify the material parameters used in our calculations. To obtain a concrete expression of the g-tensor, we adopt the following parameter set: $\eta_h=0.82,\tilde{\eta}_h=0,\gamma_h=2.62,\Delta_{\mathrm{HL}}=50\mathrm{meV}$~\cite{rimbach2025gapless,wangModelingPlanarGermanium2024}. In previous studies, the most commonly used material platform has been strained germanium heterostructures, for which the biaxial strain typically yields $\eta_h = 0.41$~\cite{abadillo-urielHoleSpinDrivingStrainInduced2023}. More recently, a new platform has been proposed, in which the hole spin qubits are confined at the interface between a bulk unstrained Ge substrate and a strained SiGe barrier that is lattice-matched to this Ge substrate~\cite{mauro2025hole}. Experiments have shown that the g-tensor components of holes become more sensitive to the electric field change in this new platform~\cite{costa2025buried,mauro2025hole,valvo2025electrically}. In our work, we adopt this material platform to realize the target device. Throughout this work, we adopt a larger value of $\eta_h = 0.82$, approximately twice that of the conventional case with a biaxial strain, which is motivated by the analysis in~\cite{valvo2025electrically}.

Based on the material parameters specified above, we next turn to the preparation and tuning of qubits within this platform. To start with, We prepare two GS2 qubits by tuning $\langle p_x^2\rangle=p_{x0}^2=\langle p_y^2\rangle=p_{y0}^2=\frac{m_0 \Delta_{\mathrm{HL}} q}{2(\lambda-\lambda')}$ for both qubits. For qubit 2, we apply an compression such that its $g_{yy}$ component is tuned to zero. More specifically, this can be achieved by tuning $\langle p_x^2\rangle = p_{x0}^2 + \delta x,\,
\langle p_y^2\rangle = p_{y0}^2 + \delta y$, where 
$\delta x / \delta y = \lambda / \lambda'$. 
We parameterize this deviation as: $\langle p_x^2\rangle = p_{x0}^2 + a\,\lambda,\,
\langle p_y^2\rangle = p_{y0}^2 + a\,\lambda'$, where the parameter $a$ determines the magnitude. As will be discussed later, the value of $a$ plays an important role in determining the voltage required to perform the g-tensor transfer. For qubit 1, by contrast, we tune its g-tensor directly by applying an appropriate voltage to the central gate.

According to the simulation results in~\cite{rimbach2025gapless}, $\langle p_x^2\rangle$ and $\langle p_y^2\rangle$ vary linearly with the central gate voltage only within a limited window of approximately $\pm 100~\mathrm{mV}$. We artificially restrict ourselves to small voltage changes to avoid significant charge occupation and orbital changes.  Our numerical simulations reveal that the voltage required to transfer $g_2$ to the target $g_2'$ is strongly correlated with the magnitude of the deformation parameter $a$. A larger $|a|$ demands a larger voltage change. Therefore, in order to minimize the required voltage change, we aim for a small $|a|$. However, $a$ cannot be chosen too small either, because an insufficient deformation would not generate the required Zeeman energy needed for the rotating wave approximation to remain valid. These constraints highlight the need for a well-balanced choice of $a$. This can for example be realized through optimized design architectures.

Assuming $\frac{\partial \langle p_{x,y}^2 \rangle}{\partial V_c}
=(0.064)^2~\mathrm{nm^{-2}/V}$, $g_1$ is obtained by applying $-100~\mathrm{mV}$ on the central gate of qubit 1 from the gapless operating point to ensure $|J_{zz}^{Q}|/|J_{\perp}|\gg 1$. The azimuthal angle and strength of the magnetic field are $\varphi_{B}=-19.31^\circ$ and $B = 0.857~\mathrm{T}$. By choosing the deformation parameter $a=3.7\times10^{-6}~\mathrm{nm^{-2}}$, the voltage required on the central gate for qubit 2 during the gate operation is $V=-102.57~\mathrm{mV}$. The exchange coupling is tuned to $J_0 = 2\pi \times35~\mathrm{MHz}$ which corresponds to the magnetic field strength $B = 0.857~\mathrm{T}$. With the chosen parameters, the three SCROFULOUS segments have durations $1.28\pi / J^{Q}_{zz}$, $2\pi / J^{Q}_{zz}$, and $1.28\pi / J^{Q}_{zz}$ respectively, resulting in a total gate time of $4.56\pi / J^{Q}_{zz}$, corresponding to a gate duration of $69~\mathrm{ns}$.

If $J^{Q}_{zz}<0$, we change the sign by exploiting the group property: $\sigma_x \sigma_z \sigma_x = -\sigma_z$. We only need to apply an additional basis rotation around the $x$-axis on qubit 1 in the qubit frame and transform it into
\begin{align}
    \mathcal{H}_{(1,1)}^{Q'}
    &= \frac{1}{2}\mu_B\, \hat{R}_x(\pi) \hat{R}_1 \mathbf{B}\cdot\hat{g}_1\boldsymbol{\sigma}_1
    + \frac{1}{2}\mu_B\, \hat{R}_2 \mathbf{B}\cdot\hat{g}_2\boldsymbol{\sigma}_2 \nonumber \\
    &+ \frac{1}{4}\,\boldsymbol{\sigma}_1 \cdot
    \left( \hat{R}_x(\pi) \hat{R}_1\,\mathcal{J}\, \hat{R}_2^{T} \right)
    \boldsymbol{\sigma}_2 \nonumber \\
    &= -\frac{1}{2}E_{Z,1}\sigma^Q_{Z,1}
    + \frac{1}{2}E_{Z,2}\sigma^Q_{Z,2}
    + \frac{1}{4}\,\boldsymbol{\sigma}_1 \cdot
    \mathcal{J}^{Q}_{\mathrm{new}}
    \boldsymbol{\sigma}_2,
\end{align}
where we find $\mathcal{J^{Q}}_{new}$, $J_{zz,new}^Q=-J_{zz}^Q$.

\section{SUDDEN GATE IMPLEMENTATION}
\label{Appendix D}
For the middle process where we tune the central gate voltage on qubit 2, the qubit frame Hamiltonian $\mathcal{H}^{Q}$ is transformed according to
\begin{align}
    \mathcal{H}^{Q} \rightarrow e^{i \frac{\theta}{2}\sigma_{X,2}} \mathcal{H}^{Q} e^{-i \frac{\theta}{2}\sigma_{X,2}}.
\end{align}
The corresponding time evolution operator is therefore given by: $e^{i \frac{\theta}{2}\sigma_{X,2}} e^{-i \mathcal{H}^{Q} t_2} e^{-i \frac{\theta}{2}\sigma_{X,2}}$ where $t_2$ denotes the evolution time of the middle process. The time evolution operator can be decomposed as
\begin{align}
    e^{i \frac{\theta}{2}\sigma_{X,2}} e^{-i \left(\varphi_1 \sigma_{Z,1} + \varphi_2 \sigma_{Z,2}\right)}e^{-i \varphi_3 \sigma_{Z,1} \sigma_{Z,2}}e^{-i \frac{\theta}{2}\sigma_{X,2}}.
\end{align}
The single-qubit phase term  $e^{-i \varphi_1 \sigma_{Z,1}}$ can be compensated either before or after the total sequence, since it commutes with $\sigma_{X,2}$. In contrast, the term $e^{-i \varphi_2 \sigma_{Z,2}}$ does not commute with $\sigma_{X,2}$, and its associated phase therefore cannot be removed by a simple additional compensation step. To address this issue, a proper synchronization of the sequence is required. Specifically, we tune the relevant parameters such that $\varphi_2=2\pi n$ with $n \in \mathrm{Z}$, rendering $\exp(-i \varphi_2 \sigma_{Z,2})$ equals to an identity operator. The condition for the phase $\varphi_2$ is given by
\begin{align}
    \varphi_2 = \frac{1}{2} E_{Z,2} t_2= \frac{\pi}{J^{Q}_{zz}}|\mu_B \hat{R}_2 \mathbf{B} \cdot\hat{g}_2|.
    \label{synchronization angle}
\end{align}

According to Eq.~\ref{synchronization angle}, the synchronization condition depends on the ratio between the magnetic field strength and the exchange coupling. Because the total gate time scales inversely with $J_{zz}^Q$, excessively small exchange couplings would lead to a long gate duration time. We therefore first choose an appropriate value of $J_0$ to ensure a practical gate time. A suitable magnetic field strength can therefore be chosen based on the required synchronization ratio.

\section{LOW-PASS FILTER}
\label{Appendix E}
\begin{figure}[t]
    \centering
    \includegraphics[width=0.47\textwidth]{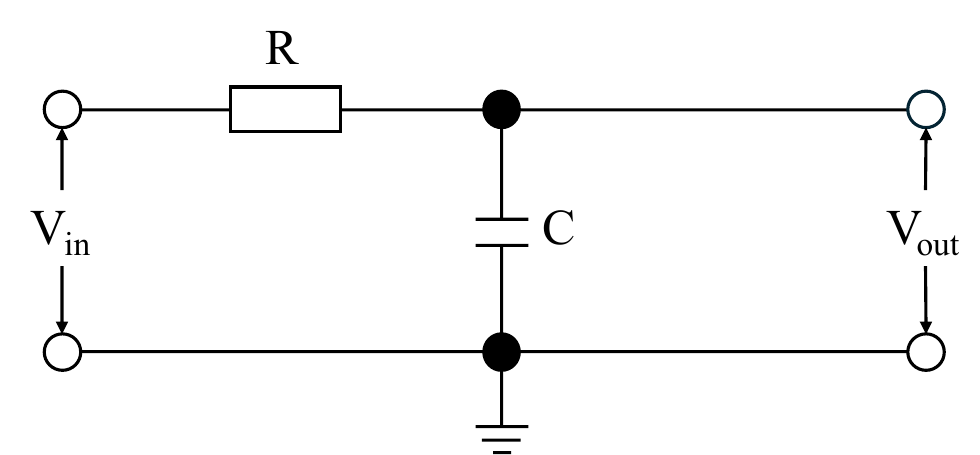}
    \caption{Schematic of a first-order RC low-pass filter used to model the finite bandwidth of the control electronics.}
    \label{fig:figure_8}
\end{figure}

We derive here demonstratively the pulse response of a simple RC low-pass filter used in the main text. Fig.~\ref{fig:figure_8} illustrates a first-order RC circuit, also known as an RC low-pass filter. Applying Kirchhoff’s voltage law to the circuit yields~\cite{irwin2010basic}
\begin{align}
  V_{\text{in}}(t) = V_R(t) + V_{\text{out}}(t). 
\end{align}
Using Ohm’s law and the capacitor relation $I(t)=C\dot V_{\text{out}}(t)$, the voltage drop across the resistor is $V_R(t)=RC \dot V_{\text{out}}(t)$. The input and output voltage therefore obey the following relation
\begin{align}
   \tau\dot V_{\text{out}}(t) + V_{\text{out}}(t) = V_{\text{in}}(t),
\end{align}
with the effective time constant $\tau = RC$.

We model the finite bandwidth of the control electronics by taking the embedded-ramp voltage pulse as the input $V_{\text{in}}(t)$ to the RC filter. The resulting output $V_{\text{out}}(t)$ thus represents the experimentally relevant, low-pass-filtered voltage waveform applied to our system. The time constant $\tau$ is chosen in the following way. Assuming the input voltage pulse to be a step function $V_{\text{in}}(t)=V_0\Theta(t)$ with $V_{\mathrm{out}}(0)=0$, the solution of $\tau \dot V_{\text{out}}(t) + V_{\text{out}}(t) = V_0$ is
\begin{align}
    V_{\text{out}}(t)=V_0\left(1-e^{-t/\tau}\right).
\end{align}
The times at which the output reaches $10\%$ and $90\%$ of its final value are defined by $V_{\text{out}}(t_{10})=0.1V_0$ and $V_{\text{out}}(t_{90})=0.9V_0$, yielding $t_{10}=-\tau\ln 0.9$ and $t_{90}=-\tau\ln 0.1$, respectively. The corresponding $10\%$--$90\%$ rise time is therefore $t_{10\text{--}90}=t_{90}-t_{10}=\tau\ln 9$. By requiring this rise time to be equal to the ramp duration $T$ we obtain $\tau = T/\ln 9\simeq T/2.2$. This choice ensures that the low-pass-filtered pulse does not deviate significantly from the original waveform while allowing the filter response to scale with the ramp duration $T$. A main deviation is the asymmetry that requires more sophisticated numerical optimization instead of simple protocols.
\begin{figure*}[t]
    \centering
    \includegraphics[width=0.85\textwidth]{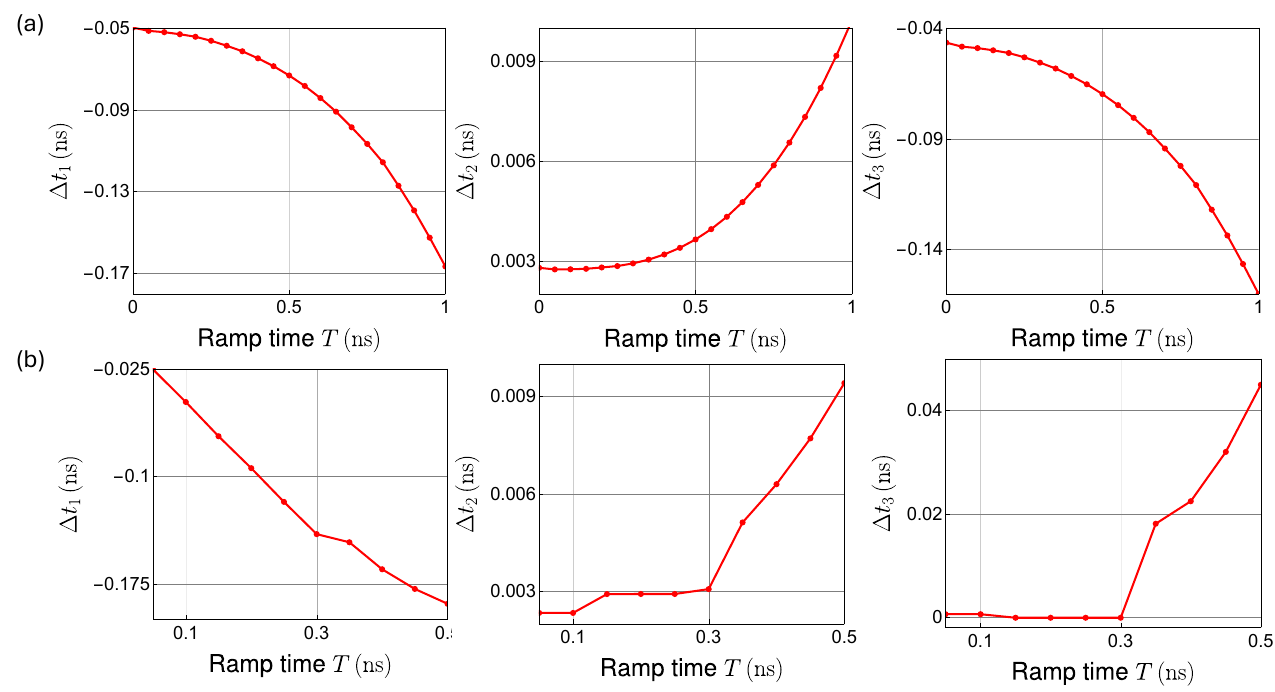}
    \caption{Numerically optimized timing corrections $\Delta t_1$, $\Delta t_2$, and $\Delta t_3$ obtained for (a) the embedded-ramp pulse and (b) the low-pass-filtered pulse.}
    \label{fig:figure_9}
\end{figure*}

\section{NUMERICAL OPTIMIZATION OF PULSE DURATION PARAMETERS}
\label{Appendix F}
In this appendix, we present the numerically obtained timing corrections $\Delta t_1$, $\Delta t_2$, and $\Delta t_3$ used to optimize the gate fidelity. Fig.~\ref{fig:figure_9}(a) and (b) demonstrate the optimized corrections for the embedded-ramp pulse and the low-pass-filtered pulse respectively. Each data point corresponds to an optimized ramp time $T$. For the low-pass-filtered pulse, the optimization starts from $T = 0.05\,\text{ns}$. To increase the convergence speed and to find continuous solutions, the values of $\Delta t_1$, $\Delta t_2$, and $\Delta t_3$ obtained at a given $T$ are used as the initial guess for the optimization subsequent optimization step. 

Interestingly, Fig.~\ref{fig:figure_9}(a) shows that at $T = 0$, corresponding to the ideal square-pulse limit, the corrections $\Delta t_1$, $\Delta t_2$, and $\Delta t_3$ do not vanish but take finite values. We have numerically verified that in the ideal square-pulse limit, using the optimized durations $t_1 + \Delta t_1$, $t_2 + \Delta t_2$, and $t_3 + \Delta t_3$ increases the gate fidelity from $0.999156$ to $0.999231$. We propose a plausible explanation: by optimizing the duration of the middle segment $t_2$, the sequence partially compensates for second-order phase shifts originating from off-diagonal terms in the effective Hamiltonian~\cite{polat2025pulse}. Together with the adjustments $\Delta t_1$ and $\Delta t_3$, this results in an enhanced gate fidelity.

\bibliography{lib}
\end{document}